\documentstyle[fleqn,12pt]{article}

\def\thalf{{\textstyle{\frac{1}{2}}}}
\def\tthird{{\textstyle{\frac{1}{3}}}}
\def\tquar{{\textstyle{\frac{1}{4}}}}
\def\pj{\hspace{-0.27cm}}
\def\fpj{\hspace{-0.7cm}}
\newcommand{\be}{\begin{equation}}
\newcommand{\ee}{\end{equation}}
\newcommand{\bq}{\begin{eqnarray}}
\newcommand{\eq}{\end{eqnarray}}
\begin{document}
\begin{titlepage}
\pagestyle{empty}
\begin{center}
\begin{large}
{{\bf  Finite Temperature Many-Body Theory with the Lipkin Model}}
\end{large}
\vskip 1cm
S.Y. Tsay Tzeng$^{a,b}$, P.J. Ellis$^{c,d}$, T.T.S. Kuo$^{b,d}$ and
E. Osnes$^{e,f} $\\
\vskip 1cm
\begin{em}{\small
$^a$ National Taipei Institute of Technology, Taipei, Taiwan,
R.O.C.}\\
\end{em}
\begin{em}{\small
$^b$Department of Physics, State University of New York at Stony Brook,
Stony Brook, NY 11794}\\
\end{em}
\begin{em}{\small
$^c$School of Physics and Astronomy, University of Minnesota, Minneapolis,
MN 55455}\\
\end{em}
\begin{em}{\small
$^d$ Institute for Nuclear Theory, University of Washington,
 Seattle, WA 98195}\\
\end{em}
\begin{em}{\small
$^e$Dept. of Physics, University of Oslo, Blindern, N-0316 Oslo 3,
Norway}\\
\end{em}
\begin{em}{\small
$^f$Theoretical Physics Institute, University of Minnesota, Minneapolis,
MN 55455}\\
\end{em}
{\bf Abstract}
\end{center}
We have compared exact numerical results for the Lipkin model at
finite temperature with Hartree-Fock theory and with the results of
including in addition the ring diagrams.  In the simplest version of the
Lipkin model the Hartree-Fock approach shows a ``phase transition" which is
absent in the exact results. For more realistic cases, Hartree-Fock provides
a very good approximation and a modest improvement is obtained by adding the
ring diagrams.
\vskip-22cm
\phantom{999}\hfill NUC-MINN-93/16-T
\end{titlepage}

\section{Introduction}

The study of hot nuclei and hot nuclear matter is of importance in heavy ion
collisions and in supernova explosions. Theoretical treatments require the
use of finite-temperature many-body theory which is inherently more difficult
than the corresponding zero-temperature formalism. Since in practical
situations the many-body
theory cannot be solved exactly, approximations are needed and it is often
difficult to know how accurate these might be. It is therefore useful to
study a model which has some realistic features, but which is sufficiently
simple to permit an exact solution, so that the accuracy of various
approximations
can be assessed. With this objective in mind we shall study here the Lipkin
model \cite{LMG65} which has been widely used at zero temperature. Rather
little work has been carried out at finite temperature with the Lipkin
model \cite{da,hat,abe} and then only for the simplest version of the
model. The principle interest in these studies was excited states and boson
expansions. Particularly striking was the demonstration that a phase
transition can occur in the Hartree-Fock (HF) solutions. We shall point out
that the exact internal energy shows a qualitatively different behavior.

In addition to HF we will
consider the correlations of the particle-particle hole-hole ring diagrams,
{\it i.e.}, the random phase approximation (RPA),
for which an expression for the grand potential has recently been given
\cite{us,nicole}. We choose these approximations since Yang, Heyer and Kuo
\cite{YHK86} found that at zero temperature HF gave a very good
approximation to the exact ground state energy and the ring series gave a
further small improvement. The final result was therefore very close to the
true energy of the system.
The present study can be regarded as the
continuation of the work of ref. \cite{YHK86} to finite temperature.
An additional motivation for studying the long
range correlations of the ring series is that in calculations with
realistic interactions it significantly
improves the location of the saturation point in nuclear matter \cite{song}.
We note in passing
that we shall not consider the particle-hole ring series since it gave
a negligible effect in the zero temperature work of ref. \cite{YHK86}.

The organization of this paper is as follows.
In subsec. 2.1 we discuss our method of obtaining exact numerical solutions for
a slightly modified version of the original Glick, Lipkin and Meshkov model
\cite{LMG65}. The approximate many-body methods with which we compare, namely
HF and RPA, are discussed in subsec. 2.2. The comparison of our
approximate and exact results is given in sec. 3 and brief concluding
remarks are presented in sec. 4.

\section{Theory}
\subsection{Exact Lipkin-Model Calculation}

The Lipkin model \cite{LMG65} consists of two single-particle levels
labelled by $\sigma=-$ and $+$, each of which has a degeneracy $p$. We write
the Hamiltonian
\bq
H&\pj=&\pj H_0+{\cal V}\;,\quad{\rm where}\nonumber\\
H_0&\pj=&\pj\thalf\xi\sum_{p\sigma}{\sigma}a_{p\sigma}^{\dagger}a_{p\sigma}\;,
\quad{\rm and}\nonumber\\
{\cal V}&\pj=&\pj\thalf V\sum_{pp'\sigma}a^{\dagger}_{p\sigma}
a^{\dagger}_{p'\sigma}a_{p'-\sigma}a_{p-\sigma}+
\thalf W\sum_{pp'\sigma}a^{\dagger}_{p\sigma}a_{p'-\sigma}^{\dagger}
a_{p'\sigma}a_{p-\sigma}\nonumber\\
&&\qquad+\thalf U\sum_{pp'\sigma}\left[a^{\dagger}_{p\sigma}
a_{p'\sigma}^{\dagger}
a_{p'-\sigma}a_{p\sigma}+a^{\dagger}_{p\sigma}a_{p'-\sigma}^{\dagger}
a_{p'\sigma}a_{p\sigma}\right]\;.
\eq
Here $H_0$ is the unperturbed  Hamiltonian with single particle energies
$\pm\thalf\xi$. The two-body interaction, ${\cal V}$, has three terms. The
interaction $V$ acts between a pair of particles with parallel spins and
changes the spins from $++$ to $--$, or {\it vice versa.} The interaction $W$
is a spin-exchange interaction and $U$, which was not present in the original
model \cite{LMG65}, flips the spin of one particle.
It is of interest to note that the interaction does not change the value of
the degeneracy labels $pp'$.

Since each particle has only two possible states, the use of the quasi-spin
formulation was suggested by Lipkin {\it et al.} \cite{LMG65}. The
quasi-spin operators obey angular momentum commutation relations and are
defined by
\be
J_z=\thalf\sum_{p\sigma}{\sigma}a^{\dagger}_{p\sigma}a_{p\sigma}
\ ,\ J_+=\sum_pa_{p+}^{\dagger}a_{p-}\ ,\ J_-=\sum_pa_{p-}^{\dagger}a_{p+}\;.
\ee
The Hamiltonian can then be compactly expressed in the form
\be
H={\xi}J_z+\thalf V(J_+^2+J_-^2)+\thalf W(J_+J_-+J_-J_+-n)
+\thalf U(J_++J_-)(n-1)\;,
\label{LIPH}
\ee
where the number operator $n=\sum_{p\sigma}a^{\dagger}_{p\sigma}
a_{p\sigma}$.
The operator $J^2=\frac{1}{2}(J_+J_-+J_-J_+)+J_z^2$ commutes with the
Hamiltonian so the Hamiltonian matrix breaks up into
submatrices of dimension $2J+1$, each associated with different values of
$J$; for a given number of particles $N$ the largest angular momentum
corresponds to $J=\thalf N$. It is straightforward to use standard angular
momentum techniques to set up  these submatrices which can then be
diagonalized. Using a label $\alpha$ to distinguish the eigenvalues $e$,
we have $e=e(N,J,\alpha)$.

Having obtained the exact eigenvalues for all $N$, we can calculate the
grand potential corresponding to the grand canonical ensemble according to
\newpage
\bq
\Omega&\pj=&\pj-\beta^{-1}\ln Z\;,\qquad {\rm where}\nonumber\\
Z&\pj=&\pj \sum_{NJ\alpha}d(N,J)e^{-\beta\{e(N,J,\alpha)-{\mu}N\}}\;.
\eq
Here $\mu$ is the chemical
potential and $\beta=T^{-1}$ is the inverse temperature.
The quantity $d(N,J)$ gives the degeneracy, {\it i.e.}, the number of times
the angular momentum $J$ occurs for a given $N$-particle system.
The physical quantities of interest, namely the mean number of particles
$\langle N\rangle$ and the internal energy $E$, can then be obtained from the
grand potential with the usual thermodynamic relations
\be
\langle N\rangle=-\frac{\partial\Omega}{\partial\mu}\quad,\quad
E=\frac{\partial}{\partial\beta}(\beta\Omega) +\mu\langle N\rangle\;.
\ee
This yields
\bq
\langle N\rangle&\pj=&\pj Z^{-1}\sum_{NJ\alpha}Nd(N,J)
e^{-\beta\{e(N,J,\alpha)-{\mu}N\}}\nonumber\\
E&\pj=&\pj Z^{-1}\sum_{NJ\alpha}e(N,J,\alpha)d(N,J)
e^{-\beta\{e(N,J,\alpha)-{\mu}N\}}\;.
\eq

\subsection{Many-Body Approximations}

Our basic many-body approach is the Hartree-Fock approximation for which
the finite temperature formalism is well known \cite{tho}. The HF single
particle equations are
\be
\epsilon_i\delta_{ik}=\langle k|H_0|i\rangle+\sum_j
\langle kj|{\cal V}|ij\rangle f_j\;,
\ee
where HF eigenvectors are used, {\it i.e.},
$|i\rangle=c_i(+)|+\rangle+c_i(-)|-\rangle$. Also the Fermi
occupation probabilities are $f_j=\left[1+\exp(\beta\tilde{\epsilon}_j)
\right]^{-1}$ with the definition $\tilde{\epsilon}_j=\epsilon_j-\mu$.
The grand potential is then
\be
\Omega_{\rm HF}=-\beta^{-1}\sum_i\ln\left(1+e^{-\beta\tilde{\epsilon}_i}
\right)-\sum_{i>j}\langle ij|{\cal V}|ij\rangle f_if_j\;,
\ee
from which, using  eq. (5), the standard relations follow
\bq
\langle N\rangle_{\rm HF}&\pj=&\pj\sum_i f_i\nonumber\\
E_{\rm HF}&\pj=&\pj\sum_i\epsilon_if_i-\sum_{i>j}\langle ij|{\cal V}|ij
\rangle f_if_j\;.
\eq

Now we also want the grand potential giving the sum of the
particle-particle hole-hole ring diagrams. This was evaluated in refs.
\cite{us,nicole}, but actually a much simpler derivation can be
given in just a few lines. It is worthwhile to present this here.
The ring series takes the form \cite{us,yang}
\bq
\Omega_{\rm ring}&\pj=&\pj\beta^{-1}\sum_{\nu}e^{i\omega_{\nu}0^+}{\rm Tr}[
F{\cal V}-\thalf(F{\cal V})^2+\tthird(F{\cal V})^3-\ldots]\nonumber\\
&\pj=&\pj\beta^{-1}\sum_{\nu}e^{i\omega_{\nu}0^+}{\rm Tr}\ln(1+F{\cal V})\;,
\eq
where the notation $F{\cal V}$ means
$F_{ij}(i\omega_{\nu})\langle ij|{\cal V}|kl\rangle$ and the
summation is over $i>j$, {\it etc.} Here the Matsubara frequency
$\omega_{\nu}=2\pi\nu T$, with $\nu$ running over all integers.
It should also be understood that a HF basis is used to evaluate the
various quantities which arise.
The pair propagator
\be
F_{ij}(i\omega_{\nu})=-\frac{Q(ij)}{i\omega_{\nu}-\tilde{\epsilon}_{ij}}\;,
\ee
where we have introduced the simplifying notation
$Q(ij)=(1-f_i)(1-f_j)-f_if_j=1-f_i-f_j$ and
$\tilde{\epsilon}_{ij}=\tilde{\epsilon}_i+\tilde{\epsilon}_j$.
Using a matrix notation, eq. (10) can be written
\be
\Omega_{\rm ring}=\beta^{-1}\sum_{\nu}e^{i\omega_{\nu}0^+}{\rm Tr}\ln
\left[\frac{(i\omega_{\nu}-\tilde{\epsilon}){\bf 1}-Q{\cal V}}{(i\omega_{\nu}
-\tilde{\epsilon}){\bf 1}}\right]\;.
\ee
The sum over $\nu$ can be performed using eq. (13) of ref. \cite{us}
with the result
\be
\Omega_{\rm ring}=\beta^{-1}{\rm Tr}\ln\left[\frac{e^{-\beta(\tilde{
\epsilon}\,{\bf 1}+Q{\cal V})}-1}{e^{-\beta\tilde{\epsilon}\,{\bf 1}}-1}
\right]\;.
\ee
Choosing a diagonal representation, {\it viz.}
\be
\sum_{k>l}\{\tilde \epsilon_{ij}\delta_{ij,kl}+Q(ij)\langle ij|{\cal V}|kl
\rangle\}\langle kl|X_n\rangle=\Delta_n\langle ij|X_n\rangle\;,\label{eqRPA2b}
\ee
and noting that Tr ln $=$ ln det, we obtain the final form
\be
\Omega_{\rm ring}=\beta^{-1}\ln\frac{\prod_n(1-e^{-\beta\Delta_n})}
{\prod_{i>j}(1-e^{-\beta\tilde\epsilon_{ij}})}\;.
\label{eqOMR}
\ee
Thus the grand potential is the difference between that obtained with
random phase approximation (RPA) bosons and that obtained with unperturbed
fermion pairs, treated as bosons. Now we should not simply add the ring and
the HF results because the first order term of eq. (10) has already
been included in the HF contribution. Thus we must subtract this, taking
\be
\Omega_{\rm ring}'=\Omega_{\rm ring}-\sum_{i>j}\langle ij|{\cal V}|ij\rangle
f_if_j\;.
\ee
Then $\Omega_{\rm total}=\Omega_{\rm HF}+\Omega_{\rm ring}'$.

A simple, but approximate, method of obtaining the thermodynamic quantities of
interest is to use the HF result (9) for the average number of particles and
take the total energy to be $E_{\rm total}=E_{\rm HF}+\Omega_{\rm ring}'$. This
ignores the effect of the derivatives of eq. (5) upon $\Omega_{\rm ring}'$
and is therefore easy to compute. We shall comment upon this approximation
later. Our aim, however, is to evaluate $\langle N\rangle_{\rm total}$ and
$E_{\rm total}$ exactly within the HF RPA formalism and this requires the
derivatives of the HF and RPA
energies and the derivative of the HF wavefunctions.

\subsubsection{Evaluation of Derivatives}

Consider a general eigenvalue equation for a matrix which may, in general, be
non-symmetric
\be
\langle \tilde{j}|H|i\rangle=E_i\delta_{ij}\quad{\rm with}\quad
\langle \tilde{j}|i\rangle=\delta_{ij}\;,
\ee
where the vectors $|\tilde{i}\rangle$ are the biorthogonal complements to
the vectors $|i\rangle$.
Denoting partial derivatives with respect to some thermodynamic variable, $x$,
by a prime, we have
\bq
&&\fpj\langle \tilde{j}'|H|i\rangle+\langle \tilde{j}|H'|i\rangle
+\langle \tilde{j}|H|i'\rangle=E_i'\delta_{ij}\;,\\
&&\fpj\langle \tilde{j}'|i\rangle+\langle \tilde{j}|i'\rangle=0\;.
\eq
In the diagonal case, $j=i$, these equations give
\be
\langle \tilde{i}|H'|i\rangle=E_i'\;,
\ee
and in the off-diagonal case, $j\ne i$,
\be
\langle \tilde{j}|H'|i\rangle=(E_i-E_j)\langle \tilde{j}|i'\rangle\;.
\ee

For the HF case we are dealing with a symmetric matrix so that
$|\tilde{i}\rangle=|i\rangle$. Then applying the above to eq. (7), we have
\be
\epsilon_i'\delta_{ik}=(\epsilon_k-\epsilon_i)\langle k|i'\rangle+
\sum\limits_j\hspace{-.21mm}\left[\langle kj'|{\cal V}|ij\rangle f_j
+\hspace{-.3mm}\langle kj|{\cal V}|ij'\rangle f_j+\hspace{-.3mm}\langle kj|
{\cal V}|ij\rangle f_j'\right]\hspace{-.3mm}.
\ee
The derivative $f_j'\equiv\frac{\partial f_j}{\partial x}=-f_j(1-f_j)
\frac{\partial}{\partial x} [\beta(\epsilon_j-\mu)]$, which involves the
unknown $\epsilon_j'$. The derivatives of the
HF wavefunctions are constrained by eq. (19) which implies
$\langle i|i'\rangle=0$ and $\langle j|i'\rangle=-\langle j'|i\rangle$.
In our case there are only two states $|i\rangle$, and labelling these
$|1\rangle$ and $|2\rangle$, we have
\be
|1'\rangle=d|2\rangle\qquad,\qquad|2'\rangle=-d|1\rangle\;,
\ee
where $d$ is a constant. Thus we have three unknowns $\epsilon_1'$,
$\epsilon_2'$ and $d$ and these can be obtained by solving the three
independant equations (22). It is also useful to note that eq. (22) yields
the relation
\be
f_i\epsilon_i'\equiv f_i\frac{\partial\epsilon_i}{\partial x}=
\frac{\partial}{\partial x} \left[\sum\limits_{i>j}
\langle ij|{\cal V}|ij\rangle f_if_j\right]\;.
\ee

We also need the derivatives of the RPA eigenvalues $\Delta_n$. Using the
vectors $\tilde{X}$ which are biorthogonal to the vectors $X$, namely
\be
\sum_{i>j}\langle\tilde{X}_m|ij\rangle\langle ij|X_n\rangle
=\delta_{mn}\;,
\ee
it follows from eqs. (14) and (20) that we can write
\be
\frac{\partial\Delta_n}{\partial x}=\sum_{i>j,k>l}
\langle\tilde{X}_n|ij\rangle\left(\frac{\partial}{\partial x}\biggl[
\tilde{\epsilon}_{ij}\delta_{ij,kl}+Q(ij)\langle ij|{\cal V}|kl\rangle\biggr]
\right)\langle kl|X_n\rangle\;.
\ee
The derivative of the quantity in square brackets involves $\epsilon_i'$,
$f_i'$ and $|i'\rangle$ which, as we have discussed, are obtained from eq.
(22).

Using eq. (5) and the grand potential $\Omega_{\rm total}$, we then have the
exact expression for the number of particles
\bq
\langle N\rangle_{\rm total}&\pj=&\pj\sum_if_i\left(1+
\frac{\partial\epsilon_i}{\partial\mu}\right)-\sum_n
\left[e^{\beta\Delta_n}-1\right]^{-1}\frac{\partial\Delta_n}{\partial\mu}
\nonumber\\
&&\qquad\qquad\qquad\qquad+\sum_{i>j}\left[e^{\beta\tilde{\epsilon}_{ij}}
-1\right]^{-1}\frac{\partial\tilde{\epsilon}_{ij}}{\partial\mu}\;.
\eq
The internal energy is given by
\bq
E_{\rm total}&\pj=&\pj\sum_if_i\left(\epsilon_i-\beta\frac{\partial\epsilon_i}
{\partial\rho}\right)+\sum_n\left[e^{\beta\Delta_n}-1\right]^{-1}\left(\Delta_n
+\beta\frac{\partial\Delta_n}{\partial\rho}\right)\nonumber\\
&&-\sum_{i>j}\left[e^{\beta\tilde{\epsilon}_{ij}}-1\right]^{-1}
\left(\tilde{\epsilon}_{ij}+\beta\frac{\partial\tilde{\epsilon}_{ij}}
{\partial\rho}\right)-2\sum_{i>j}\langle ij|{\cal V}|ij\rangle f_if_j\;,
\eq
where $\frac{\partial}{\partial\rho}\equiv\frac{\partial}{\partial\beta}-
\frac{\mu}{\beta}\frac{\partial}{\partial\mu}$.

\section{Calculation and Results}

The first step in carrying out the calculations is to compute
the two HF energies via eq. (7) which requires that both the wave
function amplitudes $c_i(\pm)$ and the
occupation probabilities $f_i$ be self-consistent. The derivatives of the
HF eigenvalues and eigenfunctions can then be obtained from eq. (22) and
the RPA eigenvalue equation (14) solved.
Now the chemical potential $\mu$ must be chosen to reproduce the
correct number of particles. Initially this is done in the HF loop
using eq. (9), but once $\Omega_{\rm ring}'$ has been computed the number of
particles must be obtained from eq. (27). This will no longer be the desired
value, so $\mu$ has to be adjusted and another iteration carried out
and so on until the correct number of particles is obtained.

For the RPA equation (14) we need to consider a five-dimensional basis,
since the results are independant of the degeneracy labels $p,p'$.
Specifically the basis is
\bq
&a^{\dagger}_{p'1}a^{\dagger}_{p1}\ ,\ a^{\dagger}_{p'1}a^{\dagger}_{p2}
\ ,\ a^{\dagger}_{p'2}a^{\dagger}_{p1}\ ,\ a^{\dagger}_{p'2}
a^{\dagger}_{p2}\;,& p\ne p'\nonumber\\
&a^{\dagger}_{p2}a^{\dagger}_{p1}&
\eq
where 1 and 2 label the HF states.
The RPA matrix actually breaks into a $4\times4$ matrix (for $p\ne p'$) and
a $1\times1$ matrix ($p=p'$).
Although we did not use it to simplify the calculations, we point out that
in actuality the $4\times4$ matrix can be split into a $1\times1$ matrix
corresponding to the linear combination
$\left(a^{\dagger}_{p'2}a^{\dagger}_{p1}-a^{\dagger}_{p'1}a^{\dagger}_{p2}
\right)$
and a $3\times3$ matrix for the orthogonal states.

We shall discuss the case where the available states are half-filled,
{\it i.e.}, $\langle N\rangle=p$, since other choices do not yield
qualitatively different results. Also for the pure HF case the levels 1 and 2
must either be completely filled or completely empty at $T=0$, so the choice
$\langle N\rangle=p$ yields a well-defined  $T=0$ limit. For the half-filled
case the pure HF result from eq. (9) requires a chemical potential
$\mu=\thalf (\epsilon_1+\epsilon_2)$, which implies
that $f_1+f_2=1$ or $Q(12)=Q(21)=0$. This means that only the
$a^{\dagger}_{p'1}a^{\dagger}_{p1}$ and $a^{\dagger}_{p'2}a^{\dagger}_{p2}$
states yield a non-zero RPA contribution. When the chemical potential is
obtained from eq. (27) which employs $\Omega_{\rm total}$, this is no longer
precisely true,
neverthelesss the contribution of the three ``$12$" states remains small.

We shall present results for the case where $p=16$ and
$\langle N\rangle=16$ so that half of the 32 available states are
filled. We have examined other values of $\langle N\rangle=p$ and found no
qualitative differences, although the approximations are quantitatively a
little less accurate for smaller numbers of particles, as one might expect.
We will take $\xi=1$, thus implicitly measuring energies in units
of $\xi$, {\it i.e.}, the quantities we discuss are dimensionless.
Further we will choose $U=W$ since no
qualitative difference is observed if they are unequal.

For our first set of calculations we chose for the parameters of the
Hamiltonian of eq. (1) $U=W=0$ and $V=-0.65$ so that we can discuss the
situation addressed in refs. \cite{da,hat,abe}. With only $V$ non-zero
analytical solutions can be obtained for the pure HF case \cite{hat,abe}.
Defining $\epsilon_1=-\epsilon_2=-\thalf\epsilon^{\rm HF}$, a solution with
$\epsilon^{\rm HF}=1$ is always possible, {\it i.e.}, in this case the HF
Hamiltonian is unchanged from the unperturbed Hamiltonian, $H_0$.
However, if it exists, the solution of
\be
\epsilon^{\rm HF}=|V|(p-1)\tanh\tquar\beta\epsilon^{\rm HF}
\ee
gives a lower energy \cite{hat,abe}. As $T$ increases, {\it i.e.}, $\beta$
decreases, the tanh decreases until the limiting case is reached where
$\epsilon^{\rm HF}=1$; for our parameter choice the critical temperature
$T_c=2.43$. This behavior is illustrated by the dashed curve in
fig. 1 where we plot the lower eigenvalue $\epsilon_1$.
The numbers on the curves indicate
the intensity $c_1(-)^2$. This starts out at 0.55 and becomes unity after
the phase transition to the unperturbed state. The corresponding internal
energy, $E_{\rm HF}$, is shown as a function of temperature in fig. 2
(dashed curve). At the phase
transition the slope becomes discontinuous and for higher temperatures the
energy arises from the unperturbed Hamiltonian, $H_0$, only. This can be
contrasted with the exact result denoted by diamonds in fig. 2 where the
curve is smooth and there is no indication of a phase transition. We
conclude that the discontinuity is an artifact of the approximation which
is employed, without physical significance. The effect of including the
ring diagrams is indicated by the solid curve in fig. 2. We see that that
they yield a modest improvement in the results at low temperatures, in
agreement with the zero-temperature results of ref. \cite{YHK86}. At
fairly high
temperatures the effect is larger and brings the calculations close to the
exact result. However, the rings yield a discontinuous curve in the ``phase
transition" region and the results are inaccurate there.
We remark in passing that a similar ``phase transition" appears for the
case $U=V=0$ with $W\ne0$ and again the exact calculations show no evidence
for such an effect.

We may note that, as $T\rightarrow\infty$,
the occupation probabilities $f_i\rightarrow\thalf$ and therefore
$Q(ij)\rightarrow0$ so that the RPA energies $\Delta_n$ of eq. (14) are just
the HF energies $\tilde{\epsilon}_{ij}$. It is straightforward to check that in
this limit $\Omega_{\rm ring}'\rightarrow0$. The internal energy is just the
HF energy which can be written
\be
E_{\rm total}=E_{\rm HF}=\tquar\sum_{i>j}\langle ij|{\cal V}|ij\rangle=
-\tquar pW\;.
\ee
Since the pair propagator vanishes ($Q(ij)\rightarrow0$) and the
particle-hole propagator also vanishes (because it is proportional to
$[(1-f_i)f_j-f_i(1-f_j)]$ which is zero in this limit) this should be an
exact result. That is, at infinite temperature only contributions of first
order in ${\cal V}$ survive. We have verified that eq. (31) agrees with
the results of our exact calculations.
For the case shown in fig. 2, eq. (31) implies that the asymptotic internal
energy is zero.

We next examine the effect of taking the relatively modest values
$U=W=-0.02$, with the same value of $V=-0.65$. The results for this case are
given in fig. 3. As compared with
fig. 2, the change in the internal energy is small and the exact results are
quite similar in the two cases. However there is a qualitative
difference  for the HF curve which now smoothly approaches the unperturbed
result for high temperatures. The unperturbed case, which arises from $H_0$ and
is the same for all the calculations we present, is represented by the
dot-dashed curve in fig. 3. Comparing this to the other curves at low
temperatures, we see that the effect of the perturbation ${\cal V}$ is very
large indeed . The corresponding HF single particle energy and
intensity in fig. 1 are similar to before except, that there is no phase
transition and that asymptotically for $T\rightarrow\infty$ the intensity
$c_1(-)^2\rightarrow0.98$ rather than unity. The effect of including the
ring diagrams here is shown by the solid curve of fig. 3. This differs from
the previous case in the region where the transition from the low to the high
temperature behavior takes place-- the curve is now smooth and the agreement
with the exact results is much better.

As a final case we take $U=W=-0.2$, these values being comparable to $V$, which
again is $-0.65$.
The results in fig. 4 show that the internal energy, $E$, is roughly doubled
at low temperature and the pure HF approximation gives very good agreement
with the exact answer. The effect of the
ring digrams is small, but they do provide even better results at high
temperatures. Reducing the rather large value of $V$ that we have used to
make it comparable to or less than $U$ and $W$ results in even less of an
effect from the ring diagrams. This is not unexpected since the
matrix element between the 11 and 22 states dominates the RPA correlations and
this is strongly influenced by $V$.
The single particle energies here (solid curve of fig. 1) are much larger
than in the previous examples and the mixing between the basis states remains
large at high temperature. In fact for $T\rightarrow\infty$ the intensity
$c_1(-)^2\rightarrow0.66$.

Finally let us discuss the approximation of taking
$E=E_{\rm HF}+\Omega_{\rm ring}'$. As $T\rightarrow0$ the quantity
$f_j(1-f_j)$ goes exponentially to zero, in which case eq. (22) indicates that
the derivatives of the HF energies and wavefunctions become zero. Thus setting
the derivative contributions in eq. (28) to zero, the $T=0$ ring contribution
to the energy is just $-(\sum_n\Delta_n-\sum_{i>j}\tilde{\epsilon}_{ij})$,
where the summation runs over those states for which $\Delta_n$ and
$\tilde{\epsilon}_{ij}\ <0$. In other words the states with energies
less (greater) than the chemical potential are filled (unfilled). The same
expression is obtained directly from $\Omega_{\rm ring}$ in eq. (15) and
this is the well-known $T=0$ result \cite{us,nicole}. Thus, at $T=0$,
the internal energy is exactly given by
$E_{\rm HF}+\Omega_{\rm ring}'$. However as the temperature increases
this becomes an approximation and it begins to deteriorate when the slope
of the internal energy curves in figs. 2--4 starts to increase. Indeed
$\Omega_{\rm ring}'$ is always negative, whereas $(E_{\rm total}-E_{\rm HF})$
is positive in some regions. In the high temperature regime
$E_{\rm HF}+\Omega_{\rm ring}'$ gives energies that are roughly halfway
between the HF and the HF+ring results. Thus the accuracy of this
approximation can only be relied upon at low temperatures.

\section{Concluding Remarks}

We have calculated the thermodynamic properties of the Lipkin model
Hamiltonian exactly and compared with approximate many-body treatments.
The case where only the $V$ (or only the $W$) term of the Hamiltonian
is non-zero is special because the Hartree-Fock single particle states
differ from the unperturbed values only for temperatures up to some
critical $T_c$. The transition between the two situations manifests itself
as a discontinuity in the slope of the calculated quantities. However no
such effect is observed in the exact calculations. Further if the other
parameters of the Hamiltonian are allowed to differ from zero, even by a
relatively small amount, this behavior of the HF theory disappears and the
calculated curves are smooth with no discontinuities in the derivatives.
For both of these reasons we conclude that the HF ``phase transition"
is an artifact which is not likely to be relevant to actual physics.

In cases where all the parameters of the Hamiltonian are non-zero,
which we think is much more likely to be representative of actual
situations, we find very good agreement between the exact and the HF
results. We found  this somewhat surprising since the differences
between the exact and the unperturbed internal energies can be large,
{\it i.e.}, the perturbation is not in any sense small. We also
examined the effects of the particle-particle ring series and found that
accuracy demands a thermodynamically correct treatment of this contribution,
{\it i.e.}, the number of particles and the internal energy should be obtained
from the thermodynamic relations (5) using the complete grand potential
$\Omega_{\rm total}$.
The ring effects yield a modest improvement of the HF result which
is mainly evident at high temperatures, although there is a small effect in
the low-$T$ regime.
Regarding the ring contribution, we make two cautionary remarks. Firstly
we have used a rather large value of the parameter $V$ so as to investigate
the situation discussed in the previous paragraph; this has the effect of
enhancing the size of the rings. Secondly in the infinite temperature limit
we have pointed out that
the ring contribution becomes zero. Nevertheless our full approximation is
remarkably accurate and it would be interesting to see what results it
yields for more realistic Hamiltonians.
\vskip1cm

Partial support from the US Department of Energy under contracts
DE-FG02-88ER40388, DE-FG02-87ER40328 and DE-FG06-90ER40561 is
gratefully acknowledged. A grant for computer time from the University of
Minnesota Supercomputer Institute is also gratefully acknowledged.

\newpage
\begin{center}
{\large{\bf Figure Captions}}
\end{center}

\noindent Figure 1.\ \ The Hartree-Fock energy of the lower state,
$\epsilon_1$, as a
function of temperature for the three different parameter sets indicated.
The numbers
indicate the intensity of the $|-\rangle$ component of the corresponding
eigenvector, {\it i.e.}, $c_1(-)^2$.

\noindent Figure 2.\ \ Comparison of the HF, the HF and ring diagram and
the exact values of the internal energy as a function of temperature. The
three cases are denoted respectively by the dashed curve, solid curve and
by diamonds. The parameters are $V=-0.65,\ U=W=0$.

\noindent Figure 3.\ \ As for fig. 2, but with parameters
$V=-0.65,\ U=W=-0.02$. Here the dot-dashed curve gives the result obtained
with the unperturbed Hamiltonian, $H_0$, only.

\noindent Figure 4.\ \ As for fig. 2, but with parameters
$V=-0.65,\ U=W=-0.2$.
\end{document}